\documentclass{PoS}

\usepackage{epsfig}
\usepackage{graphicx}

\title{Nuclear Binding Energies and NN uncertainties}

\ShortTitle{Nuclear Binding Energies and NN uncertainties}

\author{\speaker{Rodrigo Navarro Perez}%
         \thanks{Work supported by Spanish DGI
  (grant FIS2011-24149) and Junta de Andaluc{\'{\i}a} (grant FQM225).
  R.N.P. is supported by a Mexican CONACYT grant.}\\
Departamento de F\'{\i}sica At\'{o}mica, Molecular y 
  Nuclear and Instituto Carlos I de \\ F{\'\i}sica Te\'orica y Computacional. 
Universidad de Granada, E-18071 Granada, Spain.\\
        E-mail: \email{rnavarrop@ugr.es}}

\author{J. E. Amaro\\
Departamento de F\'{\i}sica At\'{o}mica, Molecular y 
  Nuclear and Instituto Carlos I de \\ F{\'\i}sica Te\'orica y Computacional. 
Universidad de Granada, E-18071 Granada, Spain.\\
        E-mail: \email{amaro@ugr.es}}

\author{E. Ruiz Arriola\\
Departamento de F\'{\i}sica At\'{o}mica, Molecular y 
  Nuclear and Instituto Carlos I de \\ F{\'\i}sica Te\'orica y Computacional. 
Universidad de Granada, E-18071 Granada, Spain.\\
        E-mail: \email{earriola@ugr.es}}

\abstract{There is an increasing interest in quantifying the
  predictive power in nuclear structure calculations. We discuss how
  both experimental and systematic errors at the NN-level can be used
  to estimate the theoretical uncertainties by rather simple means and
  without solving the full nuclear many body problem.  We emphasize
  the role of effective interactions defined by coarse graining the NN
  potential to length scales of the order of the minimal de Broglie
  wavelength probed between nucleons in nuclei. We find an {\it a priori}
  error of $\Delta B/A \sim 0.1-0.4 {\rm MeV}$ for the binding energy
  per particle throughout the periodic table for $ 2 \le A \le 208$,
  and a linear growth of the error with density for nuclear matter
  $\Delta B/A \sim 3.75 \rho_{\rm n.m.}$ and neutron matter $\Delta
  B/N \sim 3.5 \rho_{\rm n.}$. This suggests to limit the
  computational effort in solving the Nuclear Many Body Problem to
  such an accuracy.}

\FullConference{Sixth International Conference on Quarks and Nuclear Physics\\
                 April 16-20, 2012\\
                 Ecole Polytechnique, Palaiseau,  Paris}

\begin{document}

\section{Introduction}

The present contribution is based on our recent
work~\cite{NavarroPerez:2012vr}, where an old issue is adressed from a modern
perspective. The key question is, can we quantify errors in binding
energies based on our incomplete knowledge of the NN interaction ?.

The nuclear many body problem consists of diagonalizing the 
Hamiltonian $H$ with multinucleon interactions, 
\begin{eqnarray}
H = \sum_i T_i  + \sum_{i < j} V_{2,ij} + \sum_{i < j< k} V_{3,ijk}  
+ \sum_{i < j< k < l} V_{4,ijkl} + \dots 
\end{eqnarray}
where the indices run $2 \le i \le A $ and $T_i= p_i^2/(2M_N)$ is the
kinetic energy of the $i-$th nucleon, $V_{2,ij}$ the NN potential, $
V_{3,ijk} $ the NNN potential, etc. Roughly speaking one can fix $V_2$
from the deuteron and NN-scattering data, $V_3$ from triton and
nucleon-deuteron scattering, $V_4$ from the $\alpha-$particle and any
scattering process involving four nucleons, etc.

First principles calculations in Nuclear Physics have traditionally
been dominated by the idea that {\it once} the fundamental
NN-interaction is accurately known one is left with the intricacies
and complexities of the many body problem on the theoretical side.
Actually, {\it ab initio} calculations in Nuclear Physics have been
carried out as if there was complete knowledge on the elementary
NN-dynamics.  Therefore, a long-term effort has been carried out to
substantiate this assumption by continuously improving the
NN-potentials, but little attention has been paid to determine the
uncertainty in the potentials themselves.  We have recently filled
this gap by carefully analyzing different error
sources~\cite{Perez:2012qf} and using the concept of a coarse grained
potential~\cite{Perez:2011fm}.

\section{NN potentials} 

Along the years many studies have been oriented towards improving the
NN-potentials by performing $\chi^2$-fit of a Partial Wave Analysis to
the abundant available pp and np experimental data. From a purely
statistical analysis point of view, we remind that in order to
determine reliable confidence levels on the fitting parameters,
i.e. error estimates, one must have $\chi^2 /{\rm d.o.f} \lesssim
1$. Larger values of the reduced $\chi^2$ (most Bonn,Paris, Esc or
Nijm93 potentials produced $\chi^2 /{\rm d.o.f \sim 2}$) would actually
diminish the uncertainties, as there is a large penalty to change the
most likely fit parameter values, and would produce unreliable error
estimates. Thus, an error analysis of NN phase-shifts for several
partial waves became first possible when the Nijmegen
group~\cite{Stoks:1993tb} carried out a Partial Wave Analysis (PWA)
fitting about 4000 experimental np and pp data (after rejecting
further 1000 of $3\sigma$-mutually inconsistent data) with
$\chi^2/{\rm dof} \sim 1$.

In general, fits consider differential cross sections, polarization
asymmetries, etc. for a given set of energies and angles
$(E_i,\theta_i)$. While the partial wave expansion allows to evaluate
{\it any} scattering angle, the fixed energies allow to determine
phase shitfs {\it only} at those measured energies, so that the
phase-shifts themselves become independent fitting parameters.  On the
other hand, the analyticity of the S-matrix based on the
meson-exchange picture, guarantees that the phase-shitfs are smooth
functions for real positive energy.  A handy way to combine this
energy-angle information and implementing the expected smooth energy
dependence is by introducing an auxiliary potential with a set of
adjustable parameters. The bench-marking Nijmegen fit fixed the form
of the potential to incorporate charge dependence.  It contains an
energy dependent square well operating at a distance below $1.4 {\rm fm}$,
a One-Pion-Exchange (OPE) contribution starting at $1.4 {\rm fm}$,
a One-Boson-Exchange (OBE) piece  below $2-2.5 {\rm fm}$ and an
electromagnetic contribution. Energy dependence reflects retardation
in the interaction due to unobserved virtual excitations. However, it
is inconvenient to perform Nuclear structure calculations since it
requires solving a time-dependent many body problem.  At present there
are a variety of NN (energy independent) potentials fitting a large
body of scattering data with $\chi^2/{\rm dof} \sim
1$~\cite{Stoks:1993tb,Stoks:1994wp,Wiringa:1994wb,Machleidt:2000ge,Gross:2008ps}
which allow the application of conventional stationary dynamics.
Surprissingly, error estimates on the potential fitting parameters are
never provided. In~\cite{Perez:2012qf} we analyze different error
sources, in particular the short distance non-localities of the
interaction which may depend on energy, linear and angular momentum.
While in principle these non-localities are on-shell equivalent (see
e.g. Ref.~\cite{Amghar:1995av} for a proof in a $1/M_N$ expansion)
they generate quantitative differences.

\begin{figure}[ht]
\begin{center}
\epsfig{figure=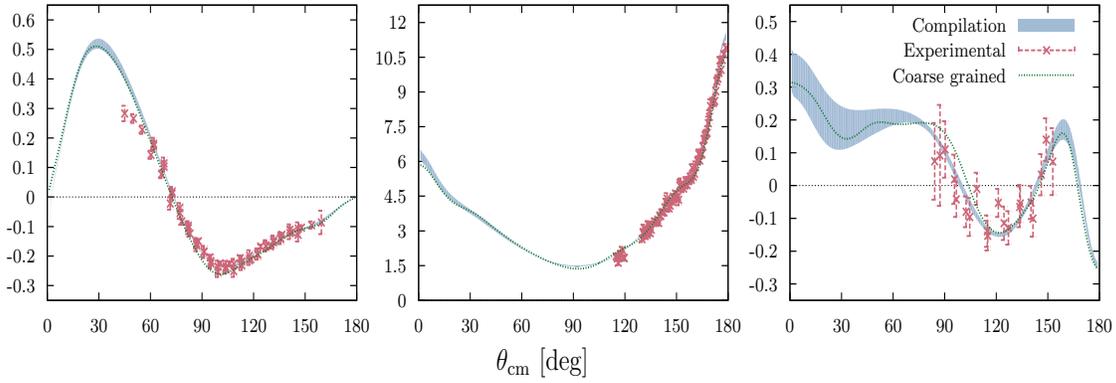,height=5cm,width=14.7cm}
\end{center}
\caption{np scattering observables for different energies in the
  laboratory system as a function of the CM angle. Left panel:
  Polarization at $E_{\rm LAB}=325 {\rm MeV}$. Middle panel:
  Differential cross section at $E_{\rm LAB}=324.1 {\rm MeV}$. Right
  panel: $D_t$ depolarization at $E_{\rm LAB}=325 {\rm MeV}$. The
  short-dashed line denotes our results by the coarse grained
  interaction~\cite{Perez:2012qf}. The band represents the compilation
  of the PWA and five high quality
  potentials~\cite{Stoks:1993tb,Wiringa:1994wb,Machleidt:2000ge,Gross:2008ps}
  which provided a $\chi^2/ {d.o.f} \lesssim 1 $.  For references for
  the experimental data see {\bf http://nn-online.org} and {\bf
    http://gwdac.phys.gwu.edu/}.}
\label{Fig1}
\end{figure}

From our reanalysis of the discrepancies in the different partial
waves and have found that the PWA statistical errors quoted in the
original Nijmegen work~\cite{Stoks:1993tb}, $\Delta \delta_{\rm PWA}$, turn out to be {\it
  smaller} than the discrepancies among the fitted phases stemming
from the different potential fits, separately claiming a $\chi^2 /
{\rm d.o.f.} \lesssim 1$ to data. In other words, 
\begin{eqnarray}
\Delta \delta_{\rm PWA} \lesssim 
\sqrt{\frac1{N-1} \sum_n (\delta_n-
  \bar \delta)^2} \, , \quad \delta_n= \delta_{\rm NijmII},\delta_{\rm Reid93},\delta_{\rm NijmI},\delta_{\rm Av18},\delta_{\rm Spec},\delta_{\rm CD Bonn}
\end{eqnarray}
Thus at $E_{\rm LAB}=350 {\rm MeV}$ in the $^1S_0$ channel one has a
s.d. of $ 0.7^{o}$ whereas $\Delta \delta_{\rm
  PWA}=0.3^{0}$. This counter-intuitive result relies not only on the
specific forms of potentials which treat the mid-- and short-range
behaviour of the interaction differently but also on the fact that the
fits are mainly done to scattering amplitudes rather than to the
phase-shifts themselves. The systematic discrepancies are vividly
illustrated by looking at Fig.~\ref{Fig1} where the spread of the
different potentials and the PWA analysis increases in the region
where no data constrain the result.

\section{Direct error estimates}

The most direct way of quantifying binding energy uncertainties would be
to undertake large scale {\it ab initio} calculations using the
different two-body potentials, say $V_2^{(i)}$ with $i=1, \dots N$,
yielding $B^{(i)}(A)$ whence a mean $\bar B(A)$ and a standard
deviation $\Delta B(A)$ can be constructed,
\begin{eqnarray}
\bar B(A) = \frac1{N} \sum_i B^{(i)}(A) \, \qquad 
\Delta B(A) = \sqrt{\frac1{N-1} \sum_i (B^{(i)}(A)-\bar B(A))^2}
\end{eqnarray}
For instance, the triton binding energy obtained by Faddeev
calculations is $ 8.00, 7.62, 7.63, 7.62, 7.72 $ and $8.50$ MeV for
the CD Bonn~\cite{Machleidt:1995km}, Nijm-II, Reid93, Nijm-I,
AV18~\cite{Friar:1993kk} and the covariant spectator
model~\cite{Gross:2008ps} respectively. This yields the combined
result $B_3 = 7.85(34) {\rm MeV}$ (exp. $B_3 = 8.4820(1) {\rm MeV}$)
i.e. $\Delta B_3/3 = 0.11 {\rm MeV}$. Note that a three-body
interaction would account for the missing $1 {\rm MeV}$-binding. 

This error estimate procedure stops beyond the $A=4$ nucleus, due to
computational and theoretical difficulties related to the form of the
potential. From an {\it ab initio} viewpoint, only Monte Carlo
calculations may go up to $A=10$ when potentials are fixed to be
$r-$dependent with a nonlocality in terms of the relative angular
momentum operator, becoming the standard Hamiltonians for {\it ab
  initio} calculations of light nuclei~\cite{Pieper:2001mp} and dense
matter~\cite{Akmal:1997ft}.  Again, a lack of error estimate of the
potential parameters makes it impossible to deduce the theoretical
errors in such calculations. On top of that, one should add the
systematic errors due to the different forms of the potential. 

Finite experimental accuracy provides also an {\it statistical} error
$\Delta V_n$ on the n-body force. In addition, unless the {\it most
  general} possible n-body forces are considered some {\it systematic}
error is introduced. On the other hand, the definition of the three-
and higher-body interaction depends on the two body potential, so any
uncertainty in the two-body interaction will {\it carry over} to the
three-body interaction. Thus, even if we fix it say in the $A=3$
system, there will always be a residual uncertainty in the $A+1=4$
calculation.  To fix ideas it is convenient to think of the n-body
potentials as random variables. They depend on some unknown parameters
$(c_1, \dots c_N)$ and will eventually be determined from $\chi^2$-fit
to some data involving n-body interactions. This would provide a
probability distribution of parameters $P(c_1, \dots c_N)$, according
to which the normalized variable $v_n= (V_n - \langle V_n
\rangle)/\sqrt{\langle (V_n - \langle V_n \rangle)^2 \rangle} $ which
has zero mean, $\langle v_n \rangle =0$, and unit variance, $\Delta
v_n=1$, can be defined.  Correlations between the two- and three-body
forces can be ignored {\it within errors} if $ (\langle v_2 v_3
\rangle)^2 \lesssim (\langle (v_2 v_3 -\langle v_2 v_3 \rangle)^2
\rangle $. Although establishing the validity of this condition
requires a thourough analysis of multinucleon forces, we will assume
this to be the case, so that 
\begin{eqnarray}
\Delta B (A)^2 = \Delta V_2^2 + \Delta V_3^2 + \Delta V_4^2 + \dots 
\end{eqnarray}
in which case, estimating the two-body uncertainty provides a lower
bound on the total uncertainty. Note that non-vanishing statistical
correlations may change this.

\section{Error estimates from Coarse grained interactions}

In our recent work~\cite{Perez:2012qf} we have shown how coarse
grained interactions~\cite{Perez:2011fm} can advantageously be used
for the purpose of error estimate.  Besides making accurate fits to
the combined NN-scattering database obtained from the PWA and the 6
high-quality potentials this representation allows to side-step the
short distance complications arising in nuclear structure
calculations. This is very much in spirit of the $V_{\rm low
  k}$-approach~\cite{Bogner:2009bt}, the Unitary Correlation Method
(UCOM)~\cite{Neff:2002nu} or the Similarity Renormalization Group
(SRG)~(see e.g. \cite{Timoteo:2011tt} and references therein), where
smooth soft-core potentials are produced after the original potential
has been evolved to the relevant scale. However, these transformations
are applied to the existing high-quality potentials and become
computationally cumbersome. We suggest instead to determine the
effective coarse grained interactions directly from fits to the
scattering data up to a given maximum energy. For light nuclei $A \le
40$ it turns out that a reasonable maximum CM momentum is about
$p_{\rm CM} \sim 200 {\rm MeV}$~\cite{Perez:2011fm} since the highest
wavelength resolution $1/p \sim 1 {\rm fm}$ is much larger than the
short range repulsion distance, $a_{\rm core} \sim 0.5 {\rm fm}$,
present in central waves. Thus, we expect mean field calculations to
be applicable.

In this way the systematic uncertainties due to the different forms of
the potential can be propagated. We have found that for doubled-closed
shell nuclei S-waves dominate the uncertainty. Using Harmonic
Oscillator Shell model wave functions we get
find~\cite{NavarroPerez:2012vr}
\begin{eqnarray}
\frac{\Delta B_{^3 {\rm H}}}{3} = 0.08(1) \, ,\quad \frac{\Delta B_{^4{\rm He}}}{4} =
0.12(1) \, , \quad 
\frac{\Delta B_{^{16}{\rm O}}}{16} = 0.28(2) \, ,  \quad \frac{\Delta B_{^{40}{\rm Ca}}}{40} = 0.34(2) {\rm MeV}
\label{eq:bind-ho}
\end{eqnarray}
the errors depending on the fitting cut-off LAB energy. While one may
reasonably doubt that variations in the binding energies can be
monitored by coarse grained interactions with harmonic oscillator wave
functions, we have found~\cite{NavarroPerez:2012vr} that this is not
the case by analyzing variations of binding energies in {\it ab
  initio} calculations induced by a relative $1\%$ change in the AV18
potential parameters~\cite{Flambaum:2007mj}.

\begin{figure}[ht]
\begin{center}
\epsfig{figure=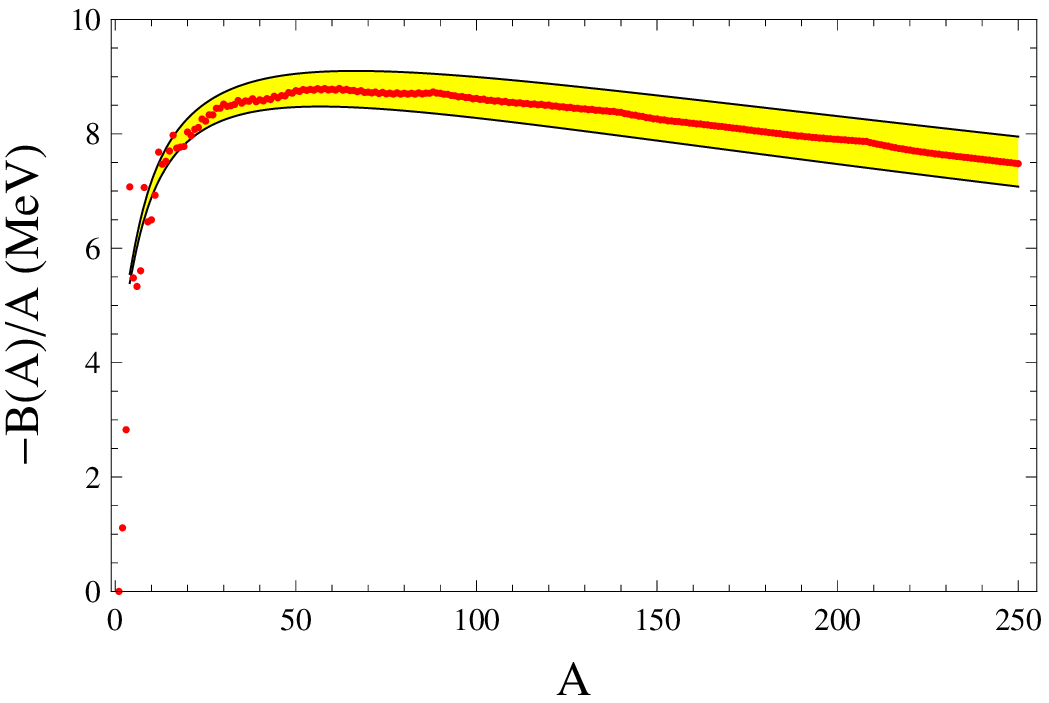,height=5cm,width=4.9cm}
\epsfig{figure=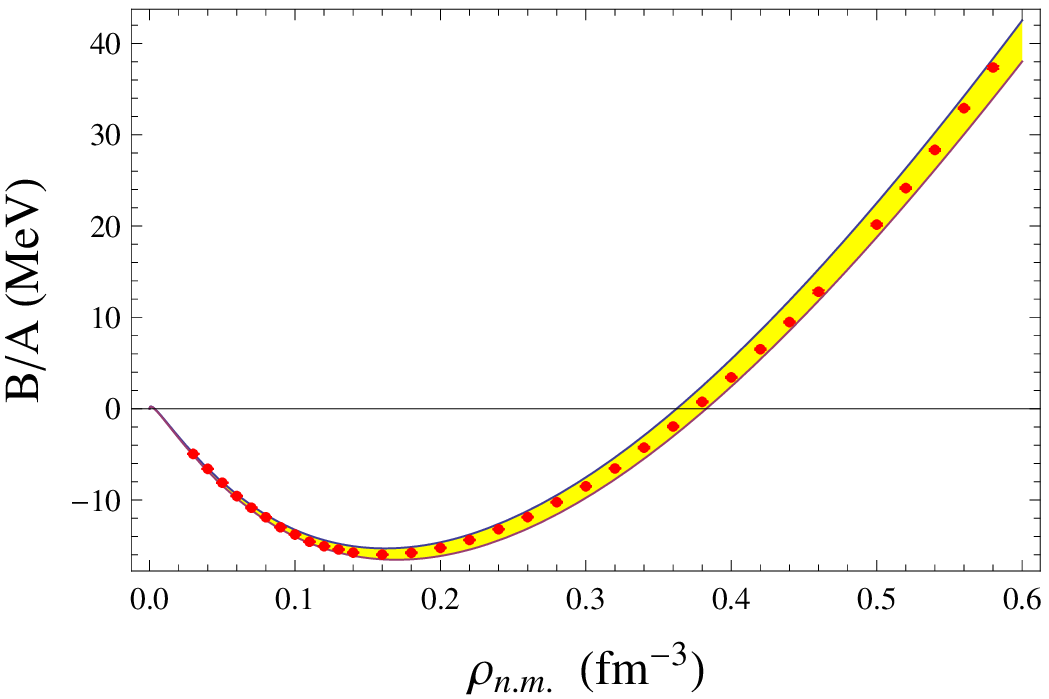,height=4.9cm,width=4.9cm}
\epsfig{figure=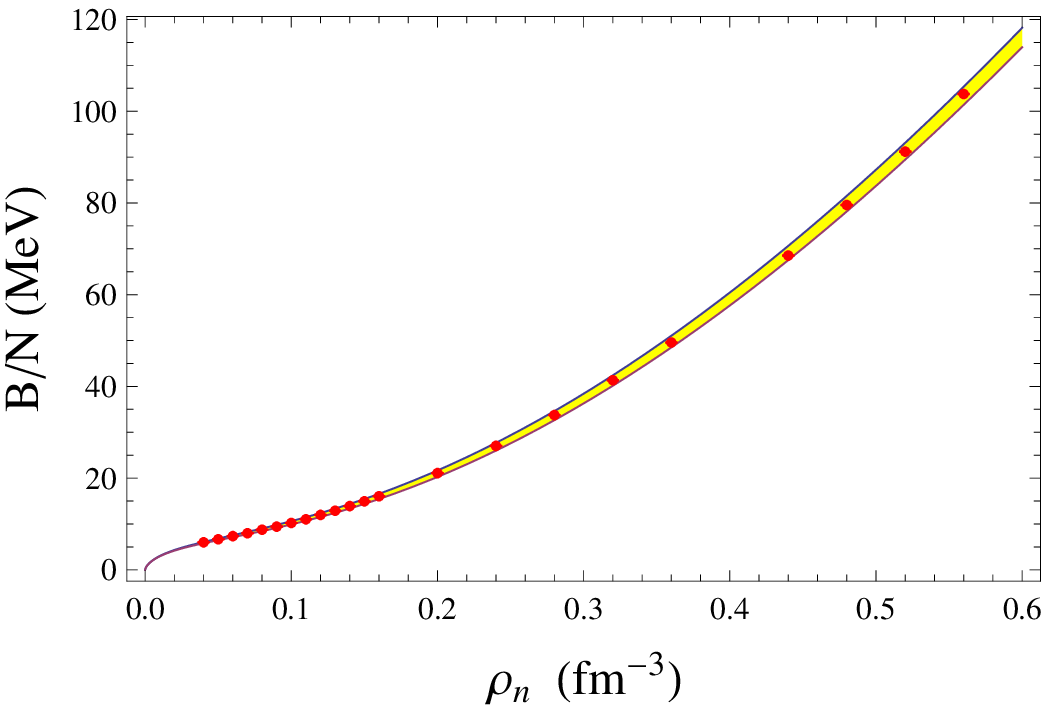,height=5cm,width=4.9cm}
\end{center}
\caption{Superposition of estimated errors based on propagating
  Nucleon-Nucleon uncertainties via coarse grained interactions. Left
  panel: Stable nuclei. Middle
  panel: Nuclear Matter Equation of State (EOS). Right Panel: Neutron Matter
  Equation of State. EOS Data are from the Green Function MonteCarlo
  Calculation of the Trento group~\cite{Gandolfi:2009nq}.}
\label{Fig2}
\end{figure}

For heavier nuclei a simple estimate of errors due to the two body
interaction uncertainty can be made by using Skyrme effective
interactions (for a review see
\cite{Bender:2003jk}) where for symmetric $N=Z=A/2$ nuclei one has
\begin{eqnarray}
\frac{\Delta B}{A} =   
\frac{3}{8 A} \Delta t_0 \, \int d^3 x \, \rho (x)^2 \,   , 
\end{eqnarray}
where $t_0 = \int d^3 x (V_{^1S_0} + V_{^3S_1})/2$ and $x_0 t_0 = \int
d^3 x (-V_{^1S_0} + V_{^3S_1})/2$.  Propagating errors we get $t_0 =
-0.92(1) {\rm GeV} {\rm fm}^3$, in agreement with the equation of state used by
the Trento group~\cite{Gandolfi:2009nq} at low densities, $t_0 \sim -
0.9(1) {\rm GeV} {\rm fm}^3 $ and with coarse graining of NN
interactions in CM momentum space down to $\Lambda\sim 0.3 {\rm GeV}$
gives a compatible value, $t_0 \sim -4
\pi^2/(M_N\Lambda)$~\cite{Arriola:2010hj}. We may implement finite
  size effects by using a Fermi-type shape for the matter density $
  \rho(r) = \rho_0/(1+e^{(r-R)/a}) $ with $R=r_0 A^{\frac13}$ and
  $r_0=1.1 {\rm fm}$ and $a=0.7 {\rm fm}$ and normalized to the total
  number of particles $A = \int d^3 x \rho(x)$ we get a result
  compatible with Eq.~(\ref{eq:bind-ho}) and depending on the value of
  $A$ for $4 \le A \le 208$. In Fig.~(\ref{Fig2}) we illustrate the
  situation by imposing our error band on the binding energy of stable nuclei.  For nuclear and neutron matter the effect grows linearly
  with the density (in ${\rm fm}^{-3}$)
\begin{eqnarray}
\frac{\Delta B_{\rm n.m.}}{A} = \frac{3}{8} \Delta t_0 \rho \sim 3.75 \rho \, , 
\qquad \frac{\Delta B_n}{N} = \frac{1}{4} \Delta \left[t_0(1-x_0) \right] \rho_n\sim 3.5 \rho_n  
\end{eqnarray}
In Fig.~\ref{Fig2} we implement our error estimates on the EOS
calculations of the Trento group~\cite{Gandolfi:2009nq}, where their
uncertainties reflect the accuracy in solving the many body problem
{\it only}.  As we see, their errors are {\it much smaller} than those
estimated here. We remind that among other effects we have neglected
possible statistical correlations between $3N$ and $2N$ forces, thus a
more thorough analysis incorporating n-body coarse grained
interactions would provide a more definite answer.

\end{document}